\documentclass[aps,preprint]{revtex4}%
\usepackage{amsfonts}
\usepackage{amsmath}
\usepackage{amssymb}
\usepackage{graphicx}%
\providecommand{\U}[1]{\protect\rule{.1in}{.1in}}
\begin{document}
\title{Switching off the reservoir through nonstationary quantum systems}
\author{L. C. C\'{e}leri$^{1}$, M. A. de Ponte$^{1}$, C. J. Villas-Boas$^{1}$ and M.
H. Y. Moussa$^{2}$.}
\affiliation{$^{1}$Departamento de F\'{\i}sica, Universidade Federal de S\~{a}o Carlos,
Caixa Postal 676, S\~{a}o Carlos, 13565-905, S\~{a}o Paulo,\textit{ }Brazil}
\affiliation{$^{2}$ Instituto de F\'{\i}sica de S\~{a}o Carlos, Universidade de S\~{a}o
Paulo, Caixa Postal 369, 13560-590 S\~{a}o Carlos, SP, Brazil }

\begin{abstract}
In this paper we demonstrate that the inevitable action of the environment can
be substantially weakened when considering appropriate nonstationary quantum
systems. Beyond protecting quantum states against decoherence, an oscillating
frequency can be engineered to make the system-reservoir coupling almost
negligible. Differently from the program for engineering reservoir and
similarly to the schemes for dynamical decoupling of open quantum systems, our
technique does not require a previous knowledge of the state to be protected.
However, differently from the previously-reported schemes for dynamical
decoupling, our technique does not rely on the availability of tailored
external pulses acting faster than the shortest time scale accessible to the
reservoir degree of freedom.

\end{abstract}

\maketitle

A great deal of attention has recently been devoted to quantum information
theory owing to its strategic position, joining up several areas of
theoretical and experimental physics. As eventually all domains of \ low
energy physics may provide potential platforms for the implementation of
quantum logic operations, efforts have been concentrated on overcoming some
sensitive problems that constitute a spectacular barrier against their
realization. These problem areas touch on both fundamental physics phenomena
--- such as decoherence and nonlocality --- and outstanding technological
issues such as individual addressing of quantum systems, separated by only a
few $\mu$m, with insignificant error \cite{Nagerl}.

As the debate around nonlocality seems to be subsiding through a set of
experimental results --- such as i) technological evidence against the
so-called loopholes \cite{Rowe}, ii) the demonstrated violation of Bell's
inequality with two-photon fringe visibilities in excess of 97\% \cite{Kwiat},
and iii) highly successful experimental quantum teleportation
\cite{Teleportation} --- the program for quantum state protection is still at
an early stage, despite all the achievements. A promising suggestion on this
subject refers to the possibility of manipulating the system-reservoir
coupling through an additional interaction between the system and a classical
ancilla. This control of decoherence through engineered reservoirs has been
theoretically implemented for atomic two-level systems, exploiting a
structured reservoir \cite{Agarwal} or mimicking a squeezed-bath interaction
\cite{Lutkenhaus}. In the domain of trapped ions, beyond a theoretical
proposition \cite{Matos}, engineered reservoirs have also been experimentally
implemented for superposed motional states of a single trapped atom
\cite{King}. Another strategy, also experimentally \ investigated
\cite{Kwiat1}, involves collective decoherence, where a composite system
interacting with a common reservoir \cite{Lidar} exhibits a decoherence-free
subspace (DFS). Whereas a common reservoir is crucial for shielding quantum
coherence in a DFS, the quantum-error correction codes QECC \cite{QECC} work,
instead, on the assumption that the decoherence process acts independently on
each of the quantum systems encoding a qubit. The issue of the physical
grounds for the assumptions behind a common or distinct reservoirs is in
detail in Ref. \cite{Mickel}.

We also mention a recent proposal for the control of coherence of a two-level
quantum system \cite{Lea}, based on random dynamical decoupling methods
\cite{Viola}. These methods resemble a previous technique to suppress
decoherence that used a tailored external driving force acting as pulses
\cite{Vitali} which, as in the present paper, was applied to a cavity-mode
superposition state. In Ref. \cite{VKL}, in a more general scope, the authors
formulated a model for decoupling a generic open quantum system from the
environmental influence also bailing out on tailored external pulses to induce
motions into the system which are faster than the shortest time scale
accessible to the reservoir degree of freedom.

In the present work we achieve the goal of Refs. \cite{Vitali,Viola,Lea,VKL},
which goes beyond the quest for quantum state protection through engineered
reservoir, from a different approach: We demonstrate, arguing from quite
general and current assumptions, that a nonstationary resonator could be
almost completely decoupled from the environment, rendering the damping factor
that characterizes the environment negligible. Note that, differently from our
proposal as well as those in Refs. \cite{Vitali,Viola,Lea,VKL}, the schemes of
engineered reservoirs require a previous knowledge of the state to be
protected. Evidently, this requirement forbids the use of engineered
reservoirs for the implementation of logic operations, making the schemes of
switching off the system-reservoir interaction more attractive. Finally, we
observe that the control of decoherence through the frequency modulation of
the system-heat--bath coupling has been proposed earlier \cite{Agarwal1}, but
as in Refs. \cite{Viola,Lea}, such control is achieved for a two-level system
instead of a cavity mode.

Assuming a nonstationary mode coupled to the environment, we get the
Hamiltonian%
\begin{equation}
H(t)=\omega(t)a^{\dagger}a+%
%TCIMACRO{\dsum \limits_{k}}%
%BeginExpansion
{\displaystyle\sum\limits_{k}}
%EndExpansion
\omega_{k}b_{k}^{\dagger}b_{k}+%
%TCIMACRO{\dsum \limits_{k}}%
%BeginExpansion
{\displaystyle\sum\limits_{k}}
%EndExpansion
\lambda_{k}(t)\left(  ab_{k}^{\dagger}+a^{\dagger}b_{k}\right)  \mathrm{,}
\label{1}%
\end{equation}
with $a^{\dagger}$ ($a$) and $b_{k}^{\dagger}$ ($b$) standing for the creation
(annihilation) operators of the nonstationary field $\omega(t)$ and the $k$th
bath mode $\omega_{k}$, respectively. Assuming the time-dependent (TD)
relation $\omega(t)=\omega_{0}-\chi\sin(\zeta t)$, the system-reservoir
couplings also turn out to be TD functions $\lambda_{k}(t)$. The simple TD
form of the free Hamiltonian $H_{0}=$ $\omega(t)a^{\dagger}a$ $+\sum
\nolimits_{k}\omega_{k}b_{k}^{\dagger}b_{k}$ enables us to describe, through
the transformation $U(t)=\exp\left(  -i\int\nolimits_{0}^{t}H_{0}(\tau
)d\tau\right)  $, the Hamiltonian in the interaction picture%
\begin{equation}
V(t)=a\Lambda^{\dagger}(t)+a^{\dagger}\Lambda(t)\mathrm{,} \label{2}%
\end{equation}
where we have defined the TD operator $\Lambda(t)=\sum_{k}\lambda_{k}%
(t)b_{k}\exp\left[  i\Delta_{k}(t)\right]  $ and parameter $\Delta
_{k}(t)=\Omega(t)-\omega_{k}t$, with $\Omega(t)=\int_{0}^{t}\omega(\tau)d\tau
$. For the case of weak system-reservoir coupling the evolution of the density
matrix of the nonstationary field, in the interaction picture and to the
second order of perturbation, is given by
\begin{equation}
\frac{d\rho(t)}{dt}=-\int_{0}^{t}dt^{\prime}\mathrm{Tr}_{R}\left[
V(t),\left[  V(t^{\prime}),\rho_{R}(0)\otimes\rho(t)\right]  \right]
\mathrm{,} \label{3}%
\end{equation}
where we have employed the usual approximation $\rho_{R}(0)\otimes\rho(t)$.
Assuming that the reservoir frequencies are very closely spaced, with spectral
density $\sigma(\mu)$, to allow the continuum summation of the coupling
strength of the resonator to the reservoir, such that $\sum_{k}\rightarrow
\left(  2\pi\right)  ^{-1}\int_{0}^{\infty}d\mu\mathbf{\sigma}\left(
\mu\right)  $, we have to solve integrals appearing in Eq. (\ref{3}), related
to correlation functions of the form%

\begin{align}
\int_{0}^{t}dt^{\prime}\left\langle \Lambda^{\dagger}(t)\Lambda(t^{\prime
})\right\rangle  &  =\operatorname*{e}\nolimits^{-i\frac{\chi}{\zeta}%
\cos(\zeta t)}\int_{0}^{t}dt^{\prime}\operatorname*{e}\nolimits^{i\frac{\chi
}{\zeta}\cos(\zeta t^{\prime})}\nonumber\\
&  \times\int_{0}^{\infty}\frac{d\mu}{2\pi}\operatorname*{e}\nolimits^{-i(\mu
-\omega_{0})(t-t^{\prime})}\sigma(\mu)N(\mu)\lambda(\mu,t)\lambda
(\mu,t^{\prime})\mathrm{,} \label{4}%
\end{align}
where the thermal average excitation of the reservoir $N\left(  \mu\right)  $
is defined by $\left\langle b^{\dagger}(\mu)b(\mu^{\prime})\right\rangle
=N(\mu)\delta(\mu-\mu^{\prime})$, while the system-reservoir coupling is
modeled as%
\begin{equation}
\lambda(\mu,t)=\lambda_{0}\frac{\xi^{2}}{\left(  \omega(t)-\mu\right)
^{2}+\xi^{2}}\mathrm{,} \label{5}%
\end{equation}
with the parameter $\xi$ accounting for the spectral sharpness around the TD
frequency of the nonstationary mode. It is quite reasonable, for the case of
weak system-reservoir coupling considered here, to assume a Lorentzian shape
for the function $\lambda(\mu,t)$, centered around the frequency $\omega(t)$.
\ Moreover, as expected, an estimate of the time average of the operator
$\Lambda(t)$ reveals that the TD system-reservoir coupling falls with
$\lambda_{0}/\left\vert \mu-\omega_{0}\right\vert $, so that the larger the
detuning, the smaller the coupling. Performing the variable transformations
$\tau=\zeta\left(  t-t^{\prime}\right)  $ and $\nu=\left(  \omega_{0}%
-\mu\right)  /\chi-\sin(\zeta t)$ in Eq. (\ref{4}) and assuming, as usual,
that $\sigma$\ and $N$\ are functions that vary slowly around the frequency
$\omega_{0}$, we obtain%
\begin{align}
\int_{0}^{t}dt^{\prime}\left\langle \Lambda^{\dagger}(t)\Lambda(t^{\prime
})\right\rangle  &  =\varkappa\kappa^{4}\chi N(\omega_{0})\int_{0}^{\zeta
t}d\tau\operatorname*{e}\nolimits^{-i\varepsilon F(\tau)}\nonumber\\
&  \times\int_{-\infty}^{a}\frac{d\nu}{2\pi}\frac{\operatorname*{e}%
\nolimits^{i\nu\varepsilon\tau}}{\left(  \nu^{2}+\kappa^{2}\right)  \left[
\left(  \nu+G(\tau)\right)  ^{2}+\kappa^{2}\right]  }\mathrm{,} \label{6}%
\end{align}
where, apart from the functions $F(\tau)=\cos\left(  \zeta t-\tau\right)
+\cos\left(  \zeta t\right)  -\tau\sin\left(  \zeta t\right)  $, $G(\tau
)=\sin\left(  \zeta t\right)  -\sin\left(  \zeta t-\tau\right)  $, and
$a=\omega_{0}/\chi-\sin\left(  \zeta t\right)  $, we have defined the
dimensionless parameters $\varkappa=$ $\Gamma_{0}/\zeta$, $\kappa=\xi/\chi$,
and $\varepsilon=\chi/\zeta$, where $\Gamma_{0}=\sigma(\omega_{0})\lambda
_{0}^{2}$ is the well-known damping rate of a stationary mode. Under the
assumption that $\chi/\omega_{0}\ll1$, the upper limit $a$ can be extended to
infinity and the corresponding integral can be evaluated analytically, leading
to the correlation function
\begin{align}
\int_{0}^{t}dt^{\prime}\left\langle \Lambda^{\dagger}(t)\Lambda(t^{\prime
})\right\rangle  &  =2N(\omega_{0})\varkappa\kappa^{4}\chi\int_{0}^{\zeta
t}d\tau\frac{\operatorname*{e}\nolimits^{i\varepsilon\left[  F(\tau)+\frac
{1}{2}\tau G(\tau)\right]  }}{G^{3}(\tau)(1+\Theta^{2})}\operatorname*{e}%
\nolimits^{-\varepsilon\kappa\tau}\nonumber\\
&  \times\left\{  G(\tau)\cos\left(  \frac{\varepsilon\tau G(\tau)}{2}\right)
+2\kappa\sin\left(  \frac{\varepsilon\tau G(\tau)}{2}\right)  \right\}
\nonumber\\
&  =N(\omega_{0})\gamma(t)\mathrm{.} \label{7}%
\end{align}
where $\Theta=2\kappa/G(\tau)$ and $\gamma(t)$ is related to an effective
time-dependent damping rate. For the sake of completeness, before analyzing
the influence of the parameters $\varkappa$, $\kappa$, and $\varepsilon$ on
the damping rate of a nonstationary mode, we compute its reduced density
operator. To this end, assuming a reservoir at absolute zero, where
$N(\omega_{0})=0$, we obtain from Eq. (\ref{3}) the master equation%
\begin{equation}
\frac{d\rho(t)}{dt}=2\operatorname{Re}\left[  \gamma(t)\right]  a\rho
(t)a^{\dagger}-\gamma^{\ast}(t)\rho(t)a^{\dagger}a-\gamma(t)a^{\dagger}%
a\rho(t)\mathrm{,} \label{8}%
\end{equation}
whose c-number version, for the normal ordered characteristic function
$\chi(\eta,\eta^{\ast},t)=Tr\left[  \rho(t)\exp(\eta a^{\dagger})\exp
(-\eta^{\ast}a)\right]  $, is given by%
\begin{equation}
\frac{\partial\chi(\eta,\eta^{\ast},t)}{\partial t}=-\gamma^{\ast}(t)\eta
\frac{\partial\chi(\eta,\eta^{\ast},t)}{\partial\eta}-\gamma(t)\eta^{\ast
}\frac{\partial\chi(\eta,\eta^{\ast},t)}{\partial\eta^{\ast}}\mathrm{.}
\label{9}%
\end{equation}

\bigskip Assuming a solution of the form $\chi(\eta,\eta^{\ast},t)=\chi
(\eta(t),\eta^{\ast}(t))$, we obtain $\eta(t)=\eta_{0}\operatorname*{e}%
\nolimits^{-\Gamma(t)/2}$, where $\eta_{0}\equiv\eta(t=0)$ and $\Gamma
(t)=\int_{0}^{t}\gamma(\tau)d\tau$ is the effective damping rate. Assuming, in
addition, that $\chi(\eta,\eta^{\ast},t)=\left.  \chi(\eta,\eta^{\ast
},t=0)\right\vert _{\eta\rightarrow\eta(t)}$, we obtain from the
Glauber-Sudarshan P-representation and the initial superposition state
$|\Psi(t=0)\rangle=\sum\nolimits_{\ell}c_{\ell}\left\vert \alpha_{0\ell
}\right\rangle $, the reduced density operator of the nonstationary mode%
\begin{equation}
\rho(t)=\mathcal{N}^{2}\sum\nolimits_{\ell\ell^{\prime}}C_{\ell\ell^{\prime}%
}(t)|\alpha_{\ell}(t)\rangle\langle\alpha_{\ell^{\prime}}(t)|\mathrm{,}
\label{10}%
\end{equation}
where $\alpha_{\ell}(t)=\alpha_{0\ell}\operatorname*{e}\nolimits^{-\gamma(t)}$
and
\begin{equation}
C_{\ell\ell^{\prime}}(t)=\exp\left\{  \left[  -\frac{1}{2}\left(  \left\vert
\alpha_{0\ell}\right\vert ^{2}+\left\vert \alpha_{0\ell^{\prime}}\right\vert
^{2}\right)  +\alpha_{0\ell^{\prime}}^{\ast}\alpha_{0\ell}\right]  \left[
1-\operatorname*{e}\nolimits^{-2\mathrm{Re}\left[  \Gamma(t)\right]  }\right]
\right\}  c_{\ell^{\prime}}^{\ast}c_{\ell}\mathrm{.} \label{11}%
\end{equation}
We note that, as expected, the decay rate turns out to be a real function even
when $\Gamma(t)$ is complex. For the particular case where the nonstationary
mode is prepared in the superposition state $|\Psi(0)\rangle=\mathcal{N}%
\left(  \left\vert \alpha_{0}\right\rangle +\left\vert -\alpha_{0}%
\right\rangle \right)  $, the function multiplying the nondiagonal elements of
the density matrix reads
\begin{equation}
C_{12}(t)=\exp\left[  -2|\alpha_{0}|^{2}\left(  1-\operatorname*{e}%
\nolimits^{-2\mathrm{Re}(\left[  \Gamma(t)\right]  }\right)  \right]
\mathrm{.} \label{12}%
\end{equation}

We now analyze the influence of the parameters $\varkappa$, $\kappa$, and
$\varepsilon$ on the effective damping rate $\Gamma(t)$ which, in its turn,
determines the decoherence time of the superposition $|\Psi(0)\rangle$, as
given by Eqs. (\ref{11}) and (\ref{12}). Starting with the parameter
$\varkappa=$ $\Gamma_{0}/\zeta$, a measure of the rate of variation of the
frequency $\zeta$, compared to the damping constant $\Gamma_{0}$, it is
evident from Eq. (\ref{7}), as expected, that the damping function $\Gamma(t)$
decreases in proportion to $\varkappa$. Otherwise, in the adiabatic regime
where $\zeta$ approaches $\Gamma_{0}$ we also expect $\Gamma(t)$ to be close
to the damping constant $\Gamma_{0}$. Regarding parameter $\kappa=\xi/\chi$,
which accounts for the range of oscillation of $\omega(t)$ compared to the
Lorentzian sharpness $\xi$, we expect the damping function to decrease as the
range of oscillation $\chi$ increases, as long as the variation rate $\zeta$
is adjusted to be significantly higher than $\Gamma_{0}$. When both parameters
$\kappa$ and $\varkappa$ are adjusted accordingly, to be significantly smaller
than unity, the system-reservoir coupling is weakened as well as the damping
function $\Gamma(t)$, consequently increasing the decoherence times of
superposition states. Differently from $\varkappa$, our expectation concerning
$\kappa$ is blurred in Eq. (\ref{7}): just as it is confirmed by the factor
$\kappa^{3}$, it is refuted by the decay function $\operatorname*{e}%
\nolimits^{-\varepsilon\kappa\tau}$ in the integral. Finally, the parameter
$\varepsilon=\chi/\zeta$ may also be defined as $\varepsilon=$ $\varkappa
/\kappa$, as long as the damping rate $\Gamma_{0}$ approximates the sharpness
$\xi$, weighting the contributions of parameters $\varkappa$ and $\kappa$. For
the same reason as $\kappa$, the role played by $\varepsilon$ in the behavior
of $\Gamma(t)$ is also blurred in Eq. (\ref{7}).

To clarify the role of the parameters $\varkappa$ and $\kappa$ in the damping
rate, in Fig. 1(a) we plot the function $C_{12}(t)$ against the scaled time
$\Gamma_{0}t$, considering the initial superposition $\left\vert
\Psi(0)\right\rangle =\mathcal{N}\left(  \left\vert \alpha_{0}\right\rangle
+\left\vert -\alpha_{0}\right\rangle \right)  $ with $\alpha_{0}=1$. The thick
solid line corresponds to the case of a stationary mode where $\omega
(t)=\omega_{0}$, prompting the expected result $\Gamma(t)=\Gamma_{0}t/2$.
Setting $\kappa=1/2$, the solid and dashed lines correspond to $\varkappa=1/2$
and $1/10$, respectively. As expected, the damping function decreases as the
rate of variation of the frequency increases. In fact, a higher rate of
variation works to hinder the system-reservoir coupling, lengthening the
response time of the system.{\large \ }With $\kappa=1/10$, the dashed-dotted
and dotted lines correspond to $\varkappa=1/2$ and $\varkappa=1/10$, showing
that the amplitude of oscillation $\chi$ is more effective in diminishing the
damping rate than the rate of variation $\zeta$. This unexpected result
reveals interesting aspects of the physics of nonstationary cavity modes:
first of all, as demonstrated below, $i)$ the time-dependence of $\omega(t)$
-- the values of the frequencies $\chi$ and $\zeta$ -- required to practically
switch off the system-reservoir coupling can be engineered through atom-field
interaction; furthermore, $ii)$ in the adiabatic regime, where $\zeta
/\omega_{0}$, $\chi/\omega_{0}\ll1$,{\large \ }the atom-field interaction is
still modelled by the Jaynes-Cummings interaction despite the nonstationary
mode \cite{Janowicz}. Consequently, all the protocols developed for the
implementation of processes in stationary modes --- for example, quantum state
or Hamiltonian engineering and logical devices --- become directly applicable
to the nonstationary mode considered here.

Figs. 1(b-f) display the damping process in the evolution of the amplitude of
the coherent state $\alpha(t)=\alpha_{0}\exp\left[  -i\Omega(t)-\Gamma
(t)\right]  $ composing the superposition $\left\vert \Psi(t)\right\rangle $.
All these figures refer to the same time interval as that used in Fig. 1(a),
thus leading to the same number of cycles coming from the rotation in phase
space, due to the factor $\operatorname{e}^{-i\Omega(t)}$. In all figures the
ratio $\omega_{0}/\Gamma_{0}=10$ is set to a fictitious scale to make clear
the spiraling of $\alpha(t)$. In Fig. 1(b), related to the thick solid line of
Fig.1(a), we observe the loss of excitation carrying the initial coherent
state to the vacuum state. In this figure we also plot, in a thick dotted
line, the evolution of the amplitude $-\alpha(t)$ of the other component of
the superposition state. Figs. 1(c-f) correspond respectively to the solid,
dashed, dashed-dotted and dotted lines of Fig. 1(a), showing a gradual
suppression of the loss of excitation which, differently from Fig. 1(b), does
not occur at a uniform rate, due to the oscillatory character -- coming from
Eq. (\ref{7}) -- of their corresponding curves in Fig. 1(a). Fig. 1(e) clearly
reveals this nonuniform character of the excitation loss through the distinct
gaps between the cycles described by the amplitude $\alpha(t)$ on its
(obstructed) way to the vacuum.

Next, considering some sensitive features in the present scheme to control
decoherence, we first address the time-dependent system-reservoir coupling
$\lambda_{k}(t)$, which can be justified through the treatment of two coupled
harmonic oscillators, one of them with time-dependent frequency. We start with
the usual coupling $\mathcal{C}X_{1}X_{2}$, where $X_{1}(t)=\mathcal{C}%
_{1}(t)(a_{1}+a_{1}^{\dagger})$ and $X_{2}=\mathcal{C}_{2}(a_{2}%
+a_{2}^{\dagger})$. Within the interaction picture and the rotating-wave
approximation, we end up with an time-dependent interaction of the form
$\mathcal{C}(t)\left(  a_{1}^{\dagger}a_{2}+a_{1}a_{2}^{\dagger}\right)  $,
similar to what had been considered in Refs. \cite{Garraway,Lea,Makri,Lo}.
Since a Lorentzian function applies whenever we have weak system-reservoir
coupling, the time-dependent function assumed in Eq. (\ref{5}) follows straightforwardly.

The most sensitive point, however, is the engineering of the nonstationary
mode whose state is to be protected. There is a great deal of literature
exploring nonstationary modes, especially in respect of Casimir effect
\cite{Dodonov}. We present below a scheme to engineer a nonstationary mode
$\omega(t)=\omega_{0}+\chi\sin(\zeta t)$ from the interaction of a driven
two-level atom (frequency $\omega_{a}$) with a stationary cavity mode
(frequency $\omega_{c}$) given by%

\begin{equation}
\mathbf{H}=\omega_{c}a^{\dagger}a+\frac{\omega_{a}}{2}\sigma_{z}+F(t)\left(
\sigma_{+}e^{-i\omega_{L}t}+\sigma_{-}e^{i\omega_{L}t}\right)  +G\left(
a\sigma_{+}+a^{\dagger}\sigma_{-}\right)  \text{,} \label{13}%
\end{equation}
where $\omega_{L}$ stands for the frequency of the classical driving field and
$G$ denotes the Rabi frequency. The atomic operators are given by $\sigma
_{z}=\left\vert e\right\rangle \left\langle e\right\vert -\left\vert
g\right\rangle \left\langle g\right\vert $, $\sigma_{+}=\left\vert
e\right\rangle \left\langle g\right\vert $, and $\sigma_{-}=\left\vert
g\right\rangle \left\langle e\right\vert $, $e$ and $g$ being the excited and
the ground states. We assume the atomic amplification modulated as
$F(t)=F_{0}\cos\left(  \zeta t/2+\phi\right)  $. In the interaction picture,
the transformed Hamiltonian is given by%

\begin{equation}
\mathbf{H}_{I}=G\left(  a\sigma_{+}e^{i\delta_{1}t}+a^{\dagger}\sigma
_{-}e^{-i\delta_{1}t}\right)  +F(t)\left(  \sigma_{+}e^{i\delta_{2}t}%
+\sigma_{-}e^{-i\delta_{2}t}\right)  \text{,} \label{14}%
\end{equation}
where $\delta_{1}=\omega_{a}-\omega_{c}$ and $\delta_{2}=\omega_{a}-\omega
_{L}$ are the atom-field and the atom-laser detunings. Next, we define
$H_{1}=G\left(  a\sigma_{+}e^{i\delta_{1}t}+a^{\dagger}\sigma_{-}%
e^{-i\delta_{1}t}\right)  $ and $H_{2}=F(t)\left(  \sigma_{+}e^{i\delta_{2}%
t}+\sigma_{-}e^{-i\delta_{2}t}\right)  $, and assume the highly off-resonance
laser amplification process, such that $\left\vert \delta_{2}\right\vert
>>F_{0}$,$\zeta$,$G$,$\left\vert \delta_{1}\right\vert $ with $G<<\left\vert
\delta_{1}\right\vert $. Under this assumption, the strongly oscillating terms
of $H_{2}$ lead, to a good approximation, to the effective Hamiltonian
\cite{James},
\begin{align}
\mathbf{H}_{eff}  &  =H_{1}-iH_{2}(t)\int H_{2}(\tau)d\tau\nonumber\\
&  =\omega_{c}a^{\dagger}a+\Omega(t)\sigma_{z}+g\left(  a\sigma_{+}%
+a^{\dagger}\sigma_{-}\right)  \text{,} \label{15}%
\end{align}
where $\Omega(t)=\omega_{a}/2+F^{2}(t)/\delta_{1}$. The diagonalization of
Hamiltonian $\mathbf{H}_{eff}$ is easily accomplished through the dressed
atomic basis $\left\{  |g,n\rangle\text{,}|e,n-1\rangle\right\}  $
\cite{Scully}. Under the usual assumption that $G^{2}n\ll\delta_{1}^{2}$, we
obtain the dispersive atom-field interaction:%
\begin{equation}
\mathcal{H}=\omega_{c}a^{\dagger}a+\Omega(t)\sigma_{z}+\Upsilon(t)a^{\dagger
}a\sigma_{z}\text{,} \label{16}%
\end{equation}
where the adjustment $\phi=\pi/4$ makes $\Upsilon(t)=\Upsilon_{1}+\Upsilon
_{2}\sin(\zeta t)$ with $\Upsilon_{1}=\left[  1-3F_{0}^{2}/2\delta_{1}%
\delta_{2}\right]  G^{2}/\delta_{1}$ and $\Upsilon_{2}=\left(  G^{2}%
/\delta_{1}\right)  \left(  F_{0}^{2}/2\delta_{1}\delta_{2}\right)  $.
Evidently, by turning off the laser we obtain the usual Stark shift $\Upsilon
a^{\dagger}a\sigma_{z}$ with $\Upsilon=G^{2}/\delta_{1}$. In a frame rotating
with the shifted atomic frequency $\Omega(t)$, obtained through the unitary
operator $U=\exp\left[  -i\widetilde{\Omega}(t)\sigma_{z}\right]  \ $with
$\widetilde{\Omega}(t)=%
%TCIMACRO{\dint }%
%BeginExpansion
{\displaystyle\int}
%EndExpansion
\Omega(t^{\prime})dt^{\prime}$, the state vector associated with the
transformed Hamiltonian $\widetilde{\mathcal{H}}=\omega_{c}a^{\dagger
}a+\Upsilon(t)a^{\dagger}a\sigma_{z}$ is given by
\begin{equation}
|\Psi\left(  t\right)  \rangle=\operatorname*{e}\nolimits^{i\widetilde{\Omega
}(t)}\left\vert g\right\rangle \left\vert \Phi_{g}\left(  t\right)
\right\rangle +\operatorname*{e}\nolimits^{-i\widetilde{\Omega}(t)}\left\vert
e\right\rangle \left\vert \Phi_{e}\left(  t\right)  \right\rangle \mathrm{,}
\label{17}%
\end{equation}
where, in the Fock basis: $\left\vert \Phi_{\ell}\left(  t\right)
\right\rangle =\sum_{n}\left\langle \ell,n\right\vert \left.  \Psi\left(
t\right)  \right\rangle \left\vert n\right\rangle $, $\ell=g,e$. Using the
orthogonality of the atomic states and Eqs. (\ref{16}) and (\ref{17}), we
obtain the uncoupled TD Schr\"{o}dinger equations%
\begin{align}
i\frac{d}{dt}|\Phi_{\ell}\left(  t\right)  \rangle &  =\widetilde{\mathcal{H}%
}_{\ell}|\Phi_{\ell}\left(  t\right)  \rangle\mathrm{,}\label{18a}\\
\widetilde{\mathcal{H}}_{\ell}  &  =\omega_{\ell}(t)a^{\dagger}a\mathrm{,}
\label{18b}%
\end{align}
where $\omega_{g}=\omega_{c}-\Upsilon(t)$ and $\omega_{e}=\omega_{c}%
+\Upsilon(t)$. Therefore, when preparing the atom in the fundamental state, we
obtain the TD frequency $\omega(t)=\omega_{0}-\chi\sin(\zeta t)$ where, from
interaction (\ref{16}), $\omega_{0}=\omega_{c}+\Upsilon_{1}$ and
$\chi=\Upsilon_{2}$. Note that the atom crosses the cavity remaining in its
ground state (due to its off-resonance interactions with both the cavity mode
and the classical field), so that there is no injection of noise coming from
the atomic decay to the engineered nonstationary cavity mode. Assuming typical
values for the parameters involved in cavity QED experiments \cite{BRH,Celso}:
$G\sim3\times10^{5}$s$^{-1}$, $\left\vert \delta_{1}\right\vert $ $\sim10^{6}%
$s$^{-1}$, $\left\vert \delta_{2}\right\vert $ $\sim10^{7}$s$^{-1}$\textbf{, }
and $\Gamma_{0}$ $\sim10^{3}$s$^{-1}$, it follows, with the intensity
$F_{0}\sim10\times G$, that $\chi\sim4\times10^{4}$s$^{-1}$ with
$\zeta\lesssim10^{6}$s$^{-1}$. Since it is reasonable to assume $\xi\sim
\Gamma_{0}$, the value $1/10$ for the parameters $\kappa$ and $\varkappa$
employed to obtain the dotted line of Fig. 1(a), is easily accomplished.

Evidently, to circumvent the difficulties introduced by the small time
interval of atom-field interaction, it would be interesting to engineer the
nonstationary mode through a sequential interaction of atoms, one by one, with
the cavity mode. The trapping of an atom inside the cavity, along the lines
suggested in Ref. \cite{Prancha}, is also a possibility to be analyzed.
Otherwise, nonstationary modes can also be achieved by other schemes as the
mechanical motion of the cavity walls \cite{Yu}, suitable for our purpose
since the frequency attainable is in the gigahertz range, or even more
sophisticated schemes where the effective motion of the walls is generated by
the excitation of a plasma in a semiconductor \cite{Braggio}%
.\textquotedblright

We have thus presented in this paper a scheme which practically switches off
the reservoir of a cavity field by engineering a suitable nonstationary mode
$\omega(t)$. Besides analyzing the physical parameters required to accomplish
this process, we also demonstrated how to engineer such a nonstationary mode
through its dispersive interaction with a driven atomic system. Evidently, the
scheme presented here for a time-dependent cavity mode applies directly to any
oscillatory system such as trapped ions, nanomechanical oscillators, and
superconducting transmission lines; it can also be extended to any
nonstationary quantum system. We believe that both techniques presented here,
to protect quantum states through nonstationary quantum systems and to
engineer such systems, can play an essential role in quantum information theory.

Beyond the information theory, we believe that the present work can directly
contribute to the field of cavity quantum electrodynamics, specifically in the
less-explored topic of the interaction of a two-level atom with a
nonstationary mode in the adiabatic regime. In fact, as the engineering of a
nonstationary mode is relatively easy to be accomplished in the adiabatic
regime --- through\ the mechanical motion of the cavity walls \cite{Yu} or
atom-field interaction, as demonstrated in this work --- typical quantum
optical phenomena may be investigated in this particular context.

\textbf{Acknowledgments}

We wish to express thanks for the support from FAPESP and CNPq, Brazilian agencies.

Fig. 1. In (a) we plot the function $C_{12}(t)$ against the scaled time
$\Gamma_{0}t$. Considering a plot of $\operatorname{Im}$($\alpha(t)$) against
$\operatorname{Re}$($\alpha(t)$), in (b-f) we observe the damping process in
the evolution of the amplitude of the coherent state $\alpha(t)$.
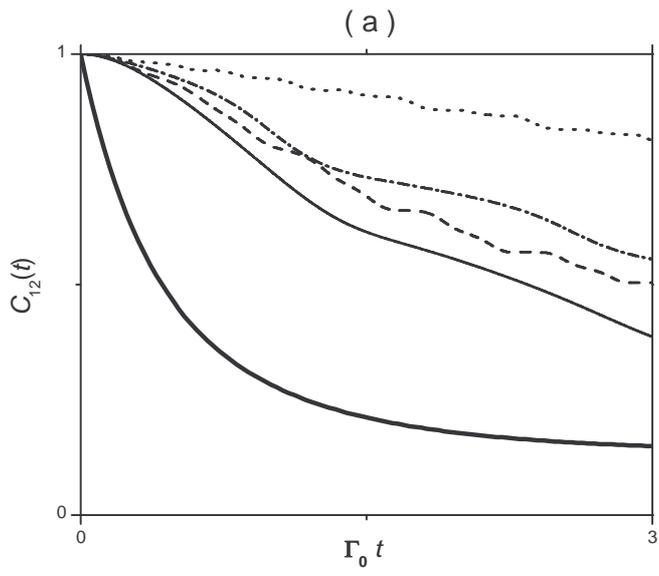
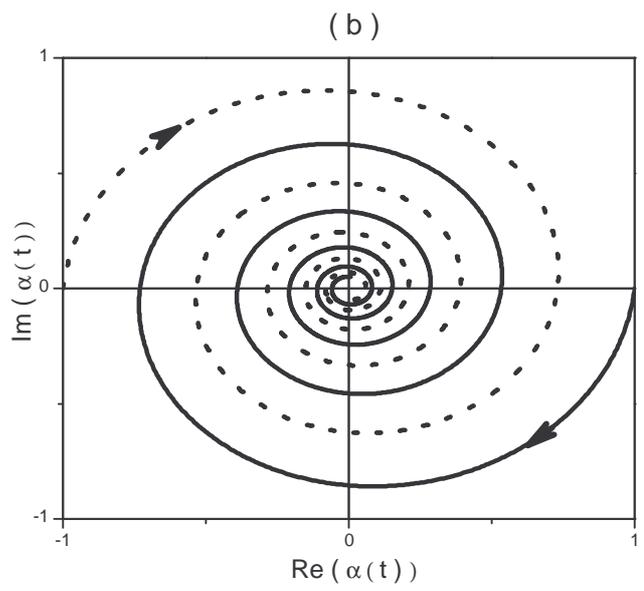
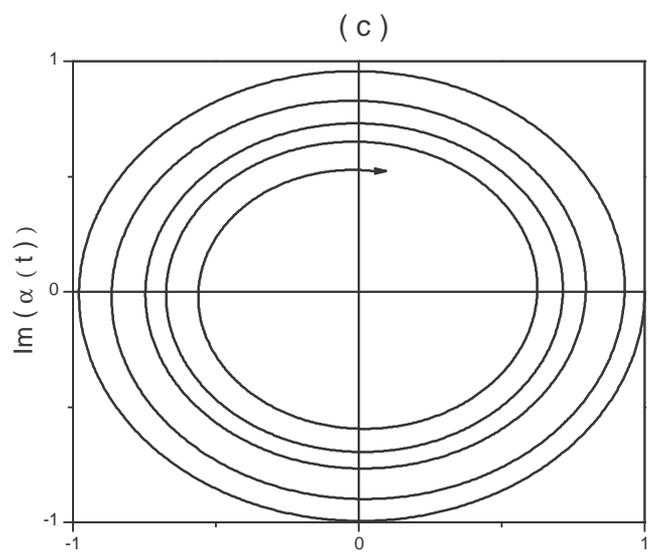
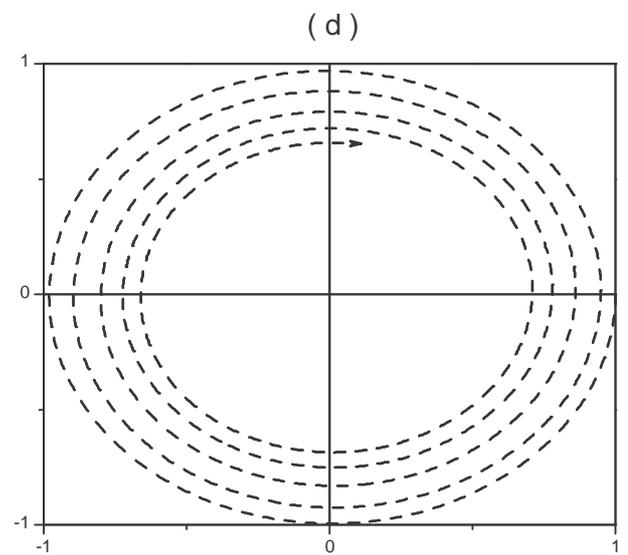
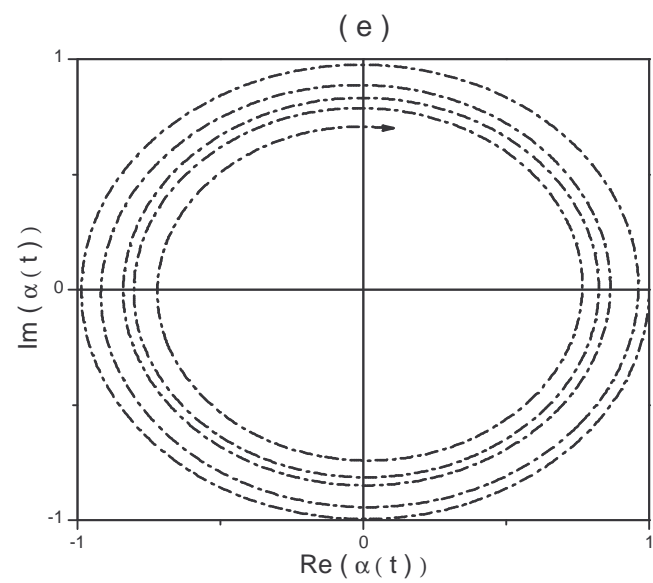
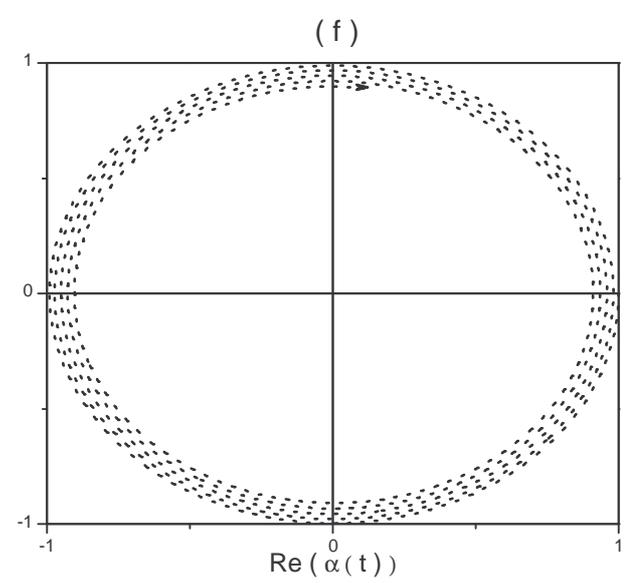
\end{document}